\documentclass[12pt, reprint, amsmath,amssymb, aps,pra,]{revtex4-1}
\usepackage[dvipdfmx]{graphicx}
\usepackage{color}
\usepackage{graphicx}% Include figure files
\usepackage{dcolumn}% Align table columns on decimal point
\usepackage{bm}% bold math
\usepackage[capitalize]{cleveref}

\begin{document}

\title{Revisiting Theoretical Analysis of Electric Dipole Moment of $^{129}$Xe}

\author{$^a$B. K. Sahoo}
\email{bijaya@prl.res.in}
\author{$^{b,c}$Nodoka Yamanaka}
\author{$^c$Kota Yanase}
\affiliation{$^a$Atomic, Molecular and Optical Physics Division, Physical Research Laboratory, Navrangpura, Ahmedabad 380009, India}
\affiliation{$^b$Kobayashi-Maskawa Institute for the Origin of Particles and the Universe, Nagoya University, Nagoya 464-8602, Japan}
\affiliation{$^c$Nishina Center for Accelerator-Based Science, RIKEN, Wako 351-0198, Japan}

\begin{abstract}
Linear response approach to the relativistic coupled-cluster (RCC) theory has been extended to estimate contributions from the parity and time-reversal violating pseudoscalar-scalar (Ps-S) and scalar-pseudoscalar (S-Ps) electron-nucleus interactions along with electric dipole moments (EDMs) of electrons ($d_e$) interacting with internal electric and magnetic fields. Random phase approximation (RPA) is also employed to produce results to compare with the earlier reported values and demonstrate importance of the non-RPA contributions arising through the RCC method. It shows that contributions from the S-Ps interactions and $d_e$ arising through the hyperfine-induced effects are very sensitive to the contributions from the high-lying virtual orbitals. Combining atomic results with the nuclear shell-model calculations, we impose constraints on the pion-nucleon coupling coefficients, and EDMs of proton and neutron. These results are further used to constrain EDMs and chromo-EDMs of up- and down-quarks by analyzing particle physics models.
\end{abstract}
\maketitle

\section{Introduction}

Searching for permanent electric dipole moments (EDMs) due to parity and time-reversal symmetry violating (P,T-odd) interactions are one of the most interesting phenomena today yet very challenging to observe in either elementary particles or composite systems \cite{ramsey,fortson}. One of the biggest cosmological mysteries in our universe is the riddle of matter-antimatter asymmetry \cite{farrar,huet,dine}. This can be explained through enough CP violating sources in the nature that are arising especially from the leptonic and semi-leptonic sources. Observations of EDMs would lead to CP violation for a wide range of sources \cite{luders}. The Standard Model (SM) of particle physics describes CP violation via a complex phase in the Cabibbo-Kobayashi-Maskawa matrix \cite{Kobayashi:1970ji}, but it cannot explain the large matter-antimatter asymmetry observed in the Universe.  Direct probes of EDMs on elementary particles are almost impossible in the next few decades as they demand energies that are beyond the reach of very large energy facilities, owing to Heisenberg's uncertainty principle, like the Large Hadron Collider (LHC) at CERN. Since EDMs of composite objects are enhanced due to electron correlation effects, atoms and molecules are used as proxies over elementary particles to fathom about CP-violating phenomena associated at the fundamental level. Although the SM predicts very small values for atomic EDMs \cite{Yamanaka:2015ncb,Yamaguchi:2020eub,Yamaguchi:2020dsy,Ema:2022yra}, the actual sizes of them could be much larger as predicted by many models beyond the SM (BSM). One would expect different types of sources of P,T-odd interactions apart from the hadronic interactions predicted by the SM within the atomic and molecular systems \cite{barr,pospelov,mjramsey,yamanaka,Chupp:2017rkp}. They can arise through the interactions among quarks, electrons and electrons and quarks. Depending on the nature of interactions, their roles become significant in a particular atomic system. Atomic EDM due to electron EDMs or P,T-odd scalar-pseudoscalar (S-Ps) electron-nucleon (e-N) interactions in diamagnetic atoms are quite small and usually neglected in the analysis. However, they can give dominant contributions to EDM of a paramagnetic system. Similarly, nuclear Schiff moment (NSM) and tensor-pseudotensor (T-Pt) e-N interactions can give significant contributions to EDM of a diamagnetic system. The former arises due to CP violating quark-gluon level interactions, such as the EDMs and chromo-EDMs of quarks. The latter is due to the T-Pt electron-quark (e-q) interaction originating from the T-Pt electron-quark interaction, which has been predicted by the leptoquark models \cite{barr1}.  

Analyzing contributions from all possible sources of P,T-odd interactions to a particular atomic system can be quite useful. Since these interactions contribute with different proportion to EDMs of various atomic systems, it would be possible to distinguish source of each type of P,T-odd interaction unambiguously by combining calculations and measurements of EDMs of a number of atomic systems. We intend to estimate contributions from as many as plausible sources of P,T-odd interactions to EDM of the $^{129}$Xe atom rigorously. As mentioned above, EDMs and chromo-EDMs of quarks as well as T-Pt e-q coefficients can be deduced from the EDM study of $^{129}$Xe atom. Compared to other diamagnetic systems, nuclear structure of $^{129}$Xe can be easily analysed theoretically. Moreover, there are three experiments underway on the measurement of EDM of $^{129}$Xe \cite{W. Heil, F. Kuchler, T. Sato}. Apart from the T-Pt e-N interactions and NSM, the other possible sources of P,T-odd interactions that can contribute to EDM of a diamagnetic system including $^{129}$Xe atom at the leading order are the pseudoscalar-scalar (Ps-S) e-N interactions, S-Ps e-N interactions and electron EDM ($d_e$) interacting with internal electric and magnetic fields \cite{Flambaum,Martensson}. Contributions from the Ps-S e-N interactions and $d_e$ interacting with the internal magnetic field can be realized at the same level of perturbation as the T-Pt e-N interactions and NSM to the EDM of the diamagnetic atoms, but their magnitudes are quite small compared to the later two interactions owing to the fact they are inversely proportional to the mass of a proton. On the other hand, the S-Ps e-N interactions and $d_e$ interacting with the internal electric field will not contribute to the EDM of diamagnetic system at the second-order of perturbation because their corresponding interaction Hamiltonians are in scalar form and the ground state of diamagnetic atoms have null angular momentum. Thus, the leading-order contributions from these interactions can arise through interactions with the magnetic dipole hyperfine ($M1_{hf}$) structure interactions. As a consequence, contributions from these interactions are also small to the EDMs of the diamagnetic atoms.

Earlier, contributions from the T-Pt e-N interactions and NSM to $^{129}$Xe were estimated rigorously by employing relativistic coupled-cluster (RCC) theory in both the linear response \cite{Y. Singh} and bi-orthogonal \cite{Sakurai} approaches, which showed results from both the approaches almost agree each other. In this work, we estimate again contributions from the T-Pt e-N interactions and NSM along with contributions from the Ps-S e-N interactions and $d_e$ interacting with nuclear magnetic field by employing the RPA and linear response RCC theory to demonstrate convergence of their values with the basis size by comparing results with the previous calculations. Then, we extend these approaches considering $M1_{hf}$ as an additional perturbation to account for the contributions from the S-Ps e-N interactions and $d_e$ interacting with the internal electric field. We find convergence of results with basis functions without and with the consideration of $M1_{hf}$ are very different, and our estimated contributions from the hyperfine induced effects differ substantially from the earlier estimations.

\section{Particle physics} \label{pph}

We can write the effective P,T-odd Lagrangian at the e-N interaction level as \cite{pospelov}
\begin{eqnarray}
\mathcal{L}_{eff}^{PT} = \mathcal{L}_e + \mathcal{L}_p + \mathcal{L}_{n} + \mathcal{L}_{\pi NN} + \mathcal{L}_{eN} ,
\end{eqnarray}
where $\mathcal{L}_e$ denotes contributions from electron EDMs, $\mathcal{L}_p$ denotes contributions from proton EDMs, $\mathcal{L}_{n}$ denotes contributions from neutron EDMs, $\mathcal{L}_{\pi NN}$ represents contributions from the pion-nucleon-nucleon ($\pi$-N-N) interactions and $\mathcal{L}_{eN} $ gives contributions from the e-N interactions. 

The relativistic expression for the EDM interaction of spin-1/2 fermion $f \, (=e , p , n)$ is given by
\begin{eqnarray}
\mathcal{L}_f = 
- \frac{i}{2} d_f \bar{\psi}_f F_{\mu \nu} \sigma^{\mu \nu} \gamma_5 \psi_f ,
\end{eqnarray}
where $F_{\mu \nu}$ is the field strength of the applied electromagnetic field, $\sigma_{\mu \nu} = \frac{i}{2}[\gamma_\mu, \gamma_\nu]$ with $\gamma$'s as the Dirac matrices, and $\psi_f$ denotes the Dirac wave function of $f$. %nucleon (electron)
The nucleon EDM is mainly generated by the EDMs of quarks at the elementary particle level. Recent lattice QCD calculations yield  \cite{Yamanaka:2018uud,Gupta:2018lvp,Alexandrou:2019brg,Horkel:2020hpi,Tsuji:2022ric,Bali:2023sdi}
\begin{eqnarray}
d_p  &\approx & 0.63 \, d_u |_{\mu = 1\, {\rm TeV}}
-0.16 \, d_d |_{\mu = 1\, {\rm TeV}} 
\end{eqnarray}
and
\begin{eqnarray}
d_n  &\approx & 0.63 \, d_d |_{\mu = 1\, {\rm TeV}}
-0.16 \, d_u |_{\mu = 1\, {\rm TeV}} ,
\end{eqnarray}
where $d_u$ and $d_d$ are the up and down quark EDMs renormalized at $\mu = 1$ TeV \cite{Yamanaka:2017mef,Degrassi:2005zd}. The extraction from experimental data is also consistent with this value \cite{Cocuzza:2023oam}, so we assign an uncertainty of 10\%.

The expression for $\mathcal{L}_e$ is given by 
\begin{eqnarray}
\mathcal{L}_e = - \frac{i}{2} d_e \bar{\psi}_e F_{\mu \nu} \sigma^{\mu \nu} \gamma_5 \psi_e  .
\end{eqnarray}

The Lagrangian for the P,T-odd $\pi$-N-N interactions that contribute significantly to the EDMs of the diamagnetic atoms is given by \cite{pospelov,Haxton:1983dq,Towner:1994qe,deVries:2020iea}
\begin{eqnarray}
\mathcal{L}_{\pi N N}  = \bar{g}_{\pi N N}^{(0)}\bar{\psi}_N \tau^i \psi_N \pi^i + \bar{g}_{\pi N N}^{(1)}
\bar{\psi}_N \psi_N \pi^0 \nonumber \\  + \bar{g}_{\pi N N}^{(2)} \big ( \bar{\psi}_N \tau^i \psi_N \pi^i - 
3 \bar{\psi}_N \tau^3 \psi_N \pi^0 \big ),
\end{eqnarray}
where the couplings $\bar{g}_{\pi N N}^{(I)}$ ($I=0,1,2$) with the superscript $i=$ 1, 2, 3 represent the isospin components. 
At the leading order, $\mathcal{L}_{\pi N N}$ is generated by the quark-gluon level CP-odd Lagrangian
\begin{eqnarray}
{\cal L}_{QCDCPV}
&=&
\Biggl(
\frac{N_q \bar \theta \alpha_s}{16\pi}
\epsilon_{\mu \nu \rho \sigma} G^{\mu \nu}_a G^{\rho \sigma}_a
\Biggl)
\nonumber\\
&&
- \sum_q^{N_q} \frac{i g_s \tilde{d}_q}{2} \bar \psi_q \sigma_{\mu \nu} G_a^{\mu \nu} t_a \gamma_5 \psi_q
\nonumber\\
&&
+
\frac{w}{6}  
f^{abc} \epsilon^{\alpha \beta \gamma \delta} G^a_{\mu \alpha } G_{\beta \gamma}^b G_{\delta}^{\ \ \mu,c}
,
\label{eq:QCDCPV}
\end{eqnarray}
where the quarks $q$ are summed over the number of active flavors $N_q$, and $G_{\mu \nu}^a$ is the field strength of the gluon with the QCD coupling $g_s$.
The first term is the so-called $\theta$-term, that we put in the parentheses because it is likely to be unphysical as shown recently  \cite{Ai:2020ptm,Nakamura:2021meh,Yamanaka:2022vdt,Yamanaka:2022bfj}. Here we write its contribution to the isoscalar CP-odd pion-nucleon interaction that was derived using the chiral perturbation theory \cite{pospelov,Chupp:2017rkp,Crewther:1979pi}
\begin{equation}
\bar g_{\pi NN}^{(0)} \approx ( 0.015\, \bar \theta ).
\end{equation}
This expression is just to let the readers know that it was believed that there were unnaturally tight constraints on $\bar \theta $ known as the strong CP problem, which can be resolved if it is unphysical. We also do not consider the Weinberg operator $w$ [last term of Eq. (\ref{eq:QCDCPV})] for which the hadron level matrix elements have large uncertainties \cite{Osamura:2022rak,Yamanaka:2020kjo,Yamanaka:2022qlu}.

The contribution of the quark chromo-EDM $\tilde{d}_q$ has also a large uncertainty, although a lot of effort has been expended in lattice QCD \cite{Abramczyk:2017oxr,Bhattacharya:2023qwf}. The leading process of $\tilde{d}_q$ contributing to the NSM is most probably the so-called vacuum alignment effect \cite{pospelov,Pospelov:2001ys}, which consists of creating a neutral pion from the vacuum by CP-odd operators.
According to chiral perturbation, this generates an isovector CP-odd $\pi$-N-N interaction \cite{Bsaisou:2014zwa,Yamanaka:2016umw,deVries:2016jox,Osamura:2022rak}
\begin{eqnarray}
&& \bar{g}_{\pi N N}^{(1)} (\tilde d_q) \nonumber\\
&\approx & -
\Biggl[
\frac{\sigma_{\pi N}}{f_\pi^2 m_\pi^2} +\frac{5 g_A^2 m_\pi}{64 \pi f_\pi^4}
\Biggr]
\frac{ f_\pi m_\pi^2 m_0^2}{2(m_u + m_d)} (\tilde{d}_u -\tilde{d}_d)
\nonumber\\
&\approx & (125 \pm 75)
\Bigl[\tilde{d}_d |_{\mu = 1\, {\rm TeV}} -\tilde{d}_u |_{\mu = 1\, {\rm TeV}}\Bigr] ,
\end{eqnarray}
where $m_\pi = 138$ MeV, $f_\pi = 93$ MeV, and $g_A=1.27$.
The quark masses are $m_u = 2.9$ MeV and $m_d=6.0$ MeV at the renormalization point $\mu =1$ GeV \cite{Yamanaka:2015ncb}. We also use the mixed condensate $m_0^2 \equiv \langle 0 | \bar \psi_q g_s \sigma_{\mu \nu} F^{\mu \nu}_a t_a \psi_q | 0 \rangle / \langle 0 | \bar q q | 0 \rangle = (0.8 \pm 0.2 )$ GeV$^2$ determined using the QCD sum rules \cite{Belyaev:1982sa,Ioffe:2005ym,Gubler:2018ctz}.
The chromo-EDM couplings are renormalized at $\mu =1$ TeV \cite{yamanaka,Degrassi:2005zd}. The uncertainty of the pion-nucleon sigma-term $\sigma_{\pi N} = (45 \pm 15) $ MeV is dominated by the systematics due to the differences between the lattice results \cite{Yamanaka:2018uud,Gupta:2021ahb,Agadjanov:2023jha,Bali:2023sdi} and phenomenological extractions \cite{Huang:2019not,Hoferichter:2023ptl}. The quoted errorbar of 60\% is a conservative one.

The leading P,T-odd Lagrangian for e-N interaction is given by \cite{pospelov}
\begin{eqnarray}
{\cal L}_{eN} &=& -\frac{G_F}{\sqrt{2}} \sum_{N} \Bigl[ 
C^{eN}_{S} \bar \psi_N \psi_N \, \bar \psi_e i \gamma^5 \psi_e \nonumber\\
&& \hspace{5em} 
+C^{eN}_{P} \bar \psi_N i\gamma^5 \psi_N \, \bar \psi_e \psi_e  \nonumber\\
&& \hspace{2em}
-\frac{1}{2}C^{eN}_{T} \varepsilon^{\mu \nu \rho \sigma} \bar \psi_N \sigma_{\mu \nu} \psi_N \, \bar \psi_e \sigma_{\rho \sigma} \psi_e  \Bigr] ,
\end{eqnarray}
where $G_F$ is the Fermi constant, $\varepsilon_{\mu \nu \alpha \beta}$ is the Levi-Civita symbol, and $\psi_{N(e)}$ denote the Dirac wave function of nucleon (electron). Here $C_S^{eN}$, $C_P^{eN}$ and $C_T^{eN}$ denote the S-Ps, Ps-S and T-Pt e-N interaction coupling constants, respectively. The above ${\cal L}_{eN}$ is generated by the  CP-odd e-q interaction, 
\begin{eqnarray}
{\cal L}_{eq} &=& -\frac{G_F}{\sqrt{2}} 
\sum_{q} \Bigl[ C^{eq}_{S} \bar \psi_q \psi_q \, \bar \psi_e i \gamma_5 \psi_e +C^{eq}_{P} \bar \psi_q i\gamma_5 \psi_q \, \bar \psi_e \psi_e \nonumber\\
&& \hspace{5em}
-\frac{1}{2}C^{eq}_{T} \varepsilon^{\mu \nu \rho \sigma} \bar \psi_q \sigma_{\mu \nu} \psi_q \, \bar \psi_e \sigma_{\rho \sigma} \psi_e \Bigr] ,
\label{eq:electronquarkinteractions}
\end{eqnarray}
at the elementary level. The relations between the CP-odd couplings are given by \cite{Yanase:2018qqq}
\begin{eqnarray}
C^{ep}_S &\approx &  11\, C^{eu}_S + 10 \, C^{ed}_S , \\
C^{en}_S &\approx & 10\, C^{eu}_S + 11\, C^{ed}_S  , \\
C^{ep}_P &\approx & 320\, C^{eu}_P  - 300\, C^{ed}_P , \\
C^{en}_P &\approx & - 300\, C^{eu}_P + 320\, C^{ed}_P , \\
C^{ep}_T &\approx & 0.63\, C^{eu}_T  -0.16\, C^{ed}_T
\end{eqnarray}
and
\begin{eqnarray}
C^{en}_T &\approx & -0.16\, C^{eu}_T +0.63\, C^{ed}_T
\end{eqnarray}
with all e-q couplings renormalized at $\mu =1$ TeV. 
The coefficients of $C^{eq}_P$ and $C^{eq}_T$ have 20\% of uncertainty, while those of $C^{eq}_S$ have 40\%, due to the systematics of the sigma-term seen above. We do not give the contributions from  the strange and heavier quarks which are affected by large errors.

\section{Nuclear physics} \label{nuph}

The NSM, $S$, is related to the P,T-odd $\pi$-N-N couplings and the nucleon EDMs as \cite{Yanase2,Yanase3}
\begin{eqnarray}
    S &=& g ( a_0 \bar{g}_{\pi N N}^{(0)}  + a_1 \bar{g}_{\pi N N}^{(1)}  + a_2 \bar{g}_{\pi N N}^{(2)} )  + b_1 d_p + b_2 d_n ,
	\label{eq:NSM coefficients} \ \ \ \
\end{eqnarray}
where $g \simeq 13.5$ is known as the strong $\pi$-N-N coupling coefficient, and $a$s and $b$s are the nuclear structure dependent coefficients. 

\begin{table}[t]
 \caption{Calculated values of $\alpha_d$ (in a.u.), $d_\mathrm{a}^{Sm}$ (in $\times 10^{-17} \frac{S}{e \ \text{fm}^3}$ e-cm), $d_\mathrm{a}^T$ (in $\times 10^{-20} \langle \sigma \rangle C_\mathrm{T}$ e-cm), $d_\mathrm{a}^{Ps}$ (in $\times 10^{-23} \langle \sigma \rangle C_\mathrm{P}$ e-cm), $d_\mathrm{a}^B$ (in $\times 10^{-4} $ e-cm), $d_\mathrm{a}^e$ (in $\times 10^{-4} $ e-cm), and $d_\mathrm{a}^{Sc}$ (in $\times 10^{-23} (C_\mathrm{S}/A)$ e-cm) from our DHF, RPA and RCCSD methods. Results from previous studies are also given including the measured value of $\alpha_d$ \cite{hohm}. We have used nuclear magnetic moment $\mu = -0.777976 \mu_N$ and nuclear spin $I=1/2$ in the estimation of hyperfine induced contributions.}
\begin{tabular}{l cccc c}
\hline \hline
Quantity & \multicolumn{4}{c}{This work}  & Others \\
 \cline{2-5} \\
         &   DHF  & RPA & RCCSD    &  Final  &         \\  
\hline \\
 $\alpha_d$ &  26.866  & 26.975  &  27.515 &  27.55(30)  & 27.815(27) \cite{hohm} \\
              &    &   &  &  & 27.782(50) \cite{Yashpal} \\
              &    &   &  &  & 27.51 \cite{sakurai} \\
              &    &   &  &  & 25.58 \cite{Fleig} \\
 $d_\mathrm{a}^{Sm}$  &  0.289  & 0.378  & 0.345 & 0.337(10) & 0.38 \cite{Dzuba} \\
               &    &   &  &  & 0.337(4) \cite{Yashpal} \\
                &    &   &  &  & 0.32 \cite{sakurai} \\
 $d_\mathrm{a}^T$ &  0.447  &  0.564  & 0.522 & 0.510(10) &  0.41 \cite{Flambaum} \\
                &   &  &  &  &  0.519  \cite{Martensson}  \\
                &   &  &  &  &  0.501(2)  \cite{Yashpal}  \\
                &    &   &  &  & 0.49 \cite{sakurai} \\
                &   &  &  &  &  0.507(48) \cite{Fleig}  \\
                &   &  &  &  &  0.57 \cite{Dzuba}  \\
 $d_\mathrm{a}^{Ps}$  & 1.287 & 1.631 & 1.504  & 1.442(25) & 1.6 \cite{Dzuba} \\
 $d_\mathrm{a}^B$   & 0.669 & 0.795 & 0.745 & 0.716(15) &  1.0 \cite{Dzuba} \\
                    &   &  &  &  &  0.869  \cite{Martensson}  \\
 $d_\mathrm{a}^e$  &  10.171  & 12.075  &  11.205  & 10.75(25) &  $-8.0$ \cite{Flambaum} \\
      &   &  &  &   &  $-9.361^{\dagger}$  \cite{Martensson}  \\
 $d_\mathrm{a}^{Sc}$ &  3.545  & 4.439  & 4.032 & 3.91(10) &  0.71(18) \cite{Fleig} \\
\hline \hline
\end{tabular}
$^{\dagger}$ Unit is changed from the original reported value using $\mu =-0.77686 \mu_N$ quoted in Ref. \cite{Martensson}.
\label{tab0}
\end{table}

\begin{table*}[t]
\caption{Convergence of the DHF values for the estimated $\alpha_d$ and EDM enhancement factors from various P,T-odd interactions in $^{129}$Xe with different sizes of basis functions which are identifies as set number (Set No.).}
\begin{tabular}{lcccc cccc}
\hline \hline
Set No. & Basis size & $\alpha_d$ & $d_\mathrm{a}^{Sm} \times 10^{-17}$ & $d_\mathrm{a}^T \times 10^{-20}$ & $d_\mathrm{a}^{Ps} \times 10^{-23}$ & $d_\mathrm{a}^B \times 10^{-4} $ & $d_\mathrm{a}^e \times 10^{-4} $ &  $d_\mathrm{a}^{Sc}\times 10^{-23}$ \\
&  & (a.u.) & (${S/(e \ \text{fm}^3)}$ e-cm) & ($\langle \sigma \rangle C_\mathrm{T}$ e-cm) & ($\langle \sigma \rangle C_\mathrm{P}$ e-cm) & e-cm & e-cm & ($(C_\mathrm{S}/A)$ e-cm) \\
 \hline \\
 I   &  $20s$, $20p$  & 4.282  & 0.289 & 0.446 & 1.286 & 0.676  & 0.640  & 0.051  \\ 
 II  &  $30s$, $30p$  &  4.282 &  0.290  & 0.447  & 1.287  &  0.675  & 8.718  & 2.017 \\
 III &  $35s$, $35p$  &  4.282 &  0.290  &  0.447 &  1.287 &  0.675  & 9.917  & 3.542 \\
 IV  &  $40s$, $40p$ &  4.282  &  0.290   & 0.447 &  1.287  & 0.675     & 9.918  & 3.547  \\
  V  &  $35s$, $35p$, $35d$  & 25.978 & 0.289 & 0.447 & 1.287 & 0.669 & 10.171 & 3.545 \\
 VI  &  $40s$, $40p$, $40d$ & 25.978  & 0.289  & 0.447 &  1.287 &  0.669  & 10.172   & 3.550  \\
 VII &  $40s$, $40p$, $40d$, $40f$, $40g$  & 26.868 & 0.289 &  0.447 & 1.287 & 0.669 & 10.172 & 3.550\\
{\bf VIII} & {\bf 35}$s$, {\bf 35}$p$, {\bf 35}$d$, {\bf 15}$f$, {\bf 15}$g$  & {\bf 26.866} & {\bf 0.289}  & {\bf 0.447} & {\bf 1.287} &  {\bf 0.669} & {\bf 10.171}  &  {\bf 3.545}  \\
  IX &  $20s$, $20p$, $20d$, $15f$, $15g$  &  26.866 & 0.289  &  0.447 & 1.287 &  0.670  & 0.651  &  0.051  \\
\hline \hline
\end{tabular}
\label{tab1}
\end{table*}

To obtain the constraints on the hadronic P,T-odd couplings, we use the results of nuclear large-scale shell model (LSSM) calculations. In this model, the nuclear effective Hamiltonian is diagonalized in an appropriate model space. For $^{129}$Xe consisting of 54 protons and 75 neutrons, we consider one major shell between the magic numbers 50 and 82 both for proton and neutron as the model space. This choice is reasonable for describing the low-energy properties of nuclei. In fact, the LSSM calculations using the effective Hamiltonians SN100PN and SNV successfully reproduce the low-energy spectra and electromagnetic moments in a wide range of nuclei. The NSM coefficients of $^{129}$Xe were reported in Refs.~\cite{Yanase1,Yanase2}. In particular, it was found that the NSM coefficient of the neutron EDM, $b_2$ in Eq.~(\ref{eq:NSM coefficients}), is apparently correlated to the nuclear magnetic moment. This demonstrates the reliability of the LSSM calculations, which reproduce with reasonable accuracy the experimental value of the magnetic moment. The KSHELL code has been utilized for the nuclear calculations~\cite{Shimizu:2019kshell}.

The NSM was evaluated as~\cite{Yanase1,Yanase2}
\begin{eqnarray}
S &=&  \bigl[ 0.002 d_p + 0.47 d_n \bigr] {\rm fm}^2 \nonumber\\
&&    + \Bigl[ - 0.038 \bar{g}_{\pi N N}^{(0)} + 0.041 \bar{g}_{\pi N N}^{(1)} + 0.082 \bar{g}_{\pi N N}^{(2)} \Bigr] g e\, {\rm fm}^3   ,
    \nonumber\\
\end{eqnarray}
where $ b_1 = -0.003 $ and $ 0.006 $ with the effective Hamiltonians SNV and SN100PN, respectively.

For completeness, we compute the nucleon spin matrix element ($\langle \sigma_{N} \rangle$) related to the T-Pt interaction in the same framework. We obtain for neutron  ($N = n$) $\langle \sigma_{n} \rangle = 0.666$ and $0.658$ by using the effective Hamiltonian SN100PN and SNV, respectively. We adopt the mean value $\langle \sigma_{n} \rangle = 0.66 $ in the following discussion. The proton ($N = p$) spin matrix element is computed as $\langle \sigma_p \rangle = 0.002$. Although this value may be model dependent, it is conclusive that the proton matrix element is orders of magnitude smaller than that of neutron.

\section{Atomic physics} \label{atph}

\subsection{Theory}

The EDM ($d_\mathrm{a}$) of an atomic system is given as the expectation value of the dipole operator $D$ in its state, the ground state $|\Psi_0 \rangle $ in this case. i.e.
 \begin{equation}
 d_\mathrm{a} = \frac{\langle \Psi_0 | D | \Psi_0 \rangle}{\langle \Psi_0 | \Psi_0 \rangle} .
\label{eq:d_a_H}
 \end{equation}
The single particle matrix element of $D$ can be found in Eq. (\ref{eqd}). Assuming that a given P,T-odd interaction in an atomic system is sufficiently smaller than the contributions from the electromagnetic interactions, we can consider up to the first-order in the P,T-odd interaction with respect to the electromagnetic interactions for the determination of atomic wave functions. This yields
\begin{eqnarray}
    | \Psi_0 \rangle \simeq | \Psi_0^{(0)} \rangle + \lambda | \Psi_0^{(1)} \rangle ,
\end{eqnarray}
where superscripts $0$ and $1$ stand for the unperturbed wave function due to electromagnetic interactions and its first-order correction due to a P,T-odd interaction Hamiltonian ($\lambda H_\mathrm{PT}$) respectively. Here $\lambda$ represents perturbative parameter of the corresponding P,T-odd interaction under consideration. In principle, all possible P,T-odd interactions need to be considered simultaneously in the determination of atomic wave function. However, it will not make any difference in the precision of the results even if we consider one type of P,T-odd interaction at a time and study their contributions subsequently in an atomic system owing to the fact that correlations among all these P,T-odd interactions are negligibly small (second-order effects are much smaller than the intended accuracy of the calculations). With the above approximation, we can express
 \begin{equation}
 	d_\mathrm{a} \simeq 2{\lambda} \frac{ \langle \Psi_0^{(0)} | D  |\Psi_0^{(1)}\rangle }{\langle \Psi_0^{(0)} | \Psi_0^{(0)} \rangle} .
	\label{edm1}
 \end{equation}
 
Considering all possible Lagrangians described in Sec. \ref{pph}, the net EDM of an atomic system can be estimated as
\begin{eqnarray} 
    d_\mathrm{a} &=&  d_\mathrm{a}^e + d_\mathrm{a}^p + d_\mathrm{a}^{n} + d_\mathrm{a}^{\pi NN } + d_\mathrm{a}^{eN} \nonumber \\
     &=& d_\mathrm{a}^e + d_\mathrm{a}^{Sm} + d_\mathrm{a}^{eN} ,
\end{eqnarray}
where superscripts denote contributions to the EDM from the respective source. We have also combined contributions from the proton EDMs, neutron EDMs, and $\pi$-N-N interactions to the net EDM contributions from the above sources and denote it as $d_\mathrm{a}^{Sm}$, which are encapsulated within the NSM ($S$).

Considering non-relativistic limit, atomic Hamiltonian accounting contributions from the electron EDM interactions is given by
\begin{eqnarray}
   H_{d_e} = 2 ic d_e \sum_k \beta_k \gamma_k^5 p_k^2  = \sum_k h_k^{d_e} , 
   \label{eqde}
\end{eqnarray}
where $c$ is the speed of light, $\beta$ and $\gamma^5$ are the Dirac matrices, and $p$ is the magnitude of the momentum of the electron. Matrix element of the single particle operator $h^{d_e}$ of $H_{d_e}$ is given by Eq. (\ref{eqde}), which shows that it is a scalar operator. As a result, Eq. (\ref{edm1}) will be zero for the closed-shell system (with total angular momentum $J=0$) when $H_{d_e}$ is considered as perturbation. To get a finite value of $d_a$ due to $H_{d_e}$ it would be necessary to consider the next leading order (third-order) interaction that can arise through the $M1_{hf}$ operator, whose matrix element is given by Eq. (\ref{m1hf}). In the presence of both P,T-odd and $M1_{hf}$ interactions, we can express an atomic wave function as 
\begin{eqnarray}
    | \Psi_0 \rangle \simeq | \Psi_0^{(0,0)} \rangle + \lambda_1 | \Psi_0^{(1,0)} \rangle + \lambda_2 | \Psi_0^{(0,1)} \rangle + \lambda_1 \lambda_2 | \Psi_0^{(1,1)} \rangle ,
    \label{eqdpt}
\end{eqnarray}
where we use $\lambda_1$ and $\lambda_2$ as perturbative parameters for $M1_{hf}$ and $H_\mathrm{PT}$ operators, respectively. Thus, the unperturbed and perturbed wave functions are denoted with two superscripts -- the first superscript counts order of $M1_{hf}$ and the second superscript counts  order of $H_\mathrm{PT}$. In these notations, we can express
\begin{eqnarray}
d_\mathrm{a}^e &=& 2 {\lambda_1} {\lambda_2} \frac{ \langle \Psi_0^{(0,0)} | D  |\Psi_0^{(1,1)}\rangle + \langle \Psi_0^{(1,0)} | D  |\Psi_0^{(0,1)}\rangle }{\langle \Psi_0^{(0,0)} | \Psi_0^{(0,0)} \rangle}  . \ \ \
\label{eq:edmdipole1}
\end{eqnarray}

 \begin{table}[t]
 \caption{Change in the DHF value for $d_\mathrm{a}^B$ (in $\times 10^{-4} $) for different values of $b$. We have used the basis set VIII and fixed $a$ as $0.523387555$ fm to carry out the analysis.}
\begin{tabular}{l cc cc}
\hline \hline
 $R$ value &  \multicolumn{4}{c}{$b$ value (in fm)} \\
 \cline {2-5} \\
 in a.u.  & $5.605$ & $5.625$ & $5.655$ & $5.695$ \\
 \hline \\
30    &  $-2.241$  &  $-2.188$ &  $-2.108$   &  $-2.001$ \\
100    & 0.581  &  1.429 &  1.365   &  1.281 \\
200    &  1.044  &  1.006 &  0.949   & 0.874 \\
500    &  0.927  &  0.721 &  0.669   &  0.600 \\
\hline \hline
\end{tabular}
\label{tab02}
\end{table}

\begin{table}[t]
\caption{The DHF values for $d_\mathrm{a}^e$ and  $d_\mathrm{a}^{Sc}$ from the basis set VIII without and after considering the nuclear magnetization distribution.}
\begin{tabular}{l cc}
\hline \hline
 Condition  &  $d_\mathrm{a}^e \times 10^{-4} $ &  $d_\mathrm{a}^{Sc}\times 10^{-23}$ \\
&   e-cm & ($(C_\mathrm{S}/A)$ e-cm) \\
 \hline \\
Without  & 11.007  & 4.624  \\
With    &  10.171  & 3.545  \\ 
\hline  \hline   
\end{tabular}
\label{tab03}
\end{table}	

Apart from contribution from $d_e$ interacting with internal electric field of an atomic system, there will also be another contribution to $d_\mathrm{a}$ because of $d_e$ interacting with the magnetic field ($B$) of the nucleus. Its interacting Hamiltonian is given by
\begin{eqnarray}
    H_B = -d_e \sum_k \gamma_k^0 B  = \sum_k h_k^B (r) .
\end{eqnarray}
The single particle matrix element of this Hamiltonian is given by Eq. (\ref{eqB}). It can contribute at the second-order perturbation to EDM as 
 \begin{equation}
 	d_\mathrm{a}^B \simeq 2{\lambda_2} \frac{ \langle \Psi_0^{(0,0)} | D  |\Psi_0^{(0,1)}\rangle }{\langle \Psi_0^{(0,0)} | \Psi_0^{(0,0)} \rangle} .
	\label{edmB}
 \end{equation}

Thus, contributions to ${d_\mathrm{a}}$ from the e-N interactions can be expressed as
\begin{eqnarray} 
 d_\mathrm{a}^{eN} = d_\mathrm{a}^P + d_\mathrm{a}^{Sc} + d_\mathrm{a}^T ,
\end{eqnarray}
where $d_\mathrm{a}^P$, $d_\mathrm{a}^{Sc}$ and $d_\mathrm{a}^T$ stand for the contributions to EDM from the Ps-S, S-Ps and T-Pt interactions, respectively.

Interaction Hamiltonian together due to $\mathcal{L}_{\pi N N}$, $\mathcal{L}_p$ and $\mathcal{L}_{n}$ for the atom with nuclear spin $I=1/2$ like $^{129}$Xe can be given approximately by \cite{V. V. Flambaum}
\begin{eqnarray}
	H_\mathrm{int}^\mathrm{NSM} &=& \sum_k \frac{3 (\bm{S} \cdot \bm{r})_k }{B} \rho_\mathrm{nuc}(r) \nonumber \\
           &=& \sum_k h_k^{NSM}(r) ,
	\label{eq:NSM}
\end{eqnarray}
where $\rho_\mathrm{nuc}(r)$ is the nuclear charge density distribution function, $\bm{S} = S \frac{\bm{I}}{I}$ is the NSM and $B = \int^{\infty}_{0}dr r^4 \rho_\mathrm{nuc}(r)$. The matrix element of $h_k^{NSM}(r)$ is given by Eq. (\ref{eqnsm}). $H_\mathrm{int}^\mathrm{NSM}$ can contribute at the second-order perturbation to EDM as 
 \begin{equation}
 	d_\mathrm{a}^{Sm} \simeq 2{\lambda_2} \frac{ \langle \Psi_0^{(0,0)} | D  |\Psi_0^{(0,1)}\rangle }{\langle \Psi_0^{(0,0)} | \Psi_0^{(0,0)} \rangle} .
	\label{edmSm}
 \end{equation}

\begin{figure}[t]
%\centering
\includegraphics[width=8.5cm,height=3.5cm]{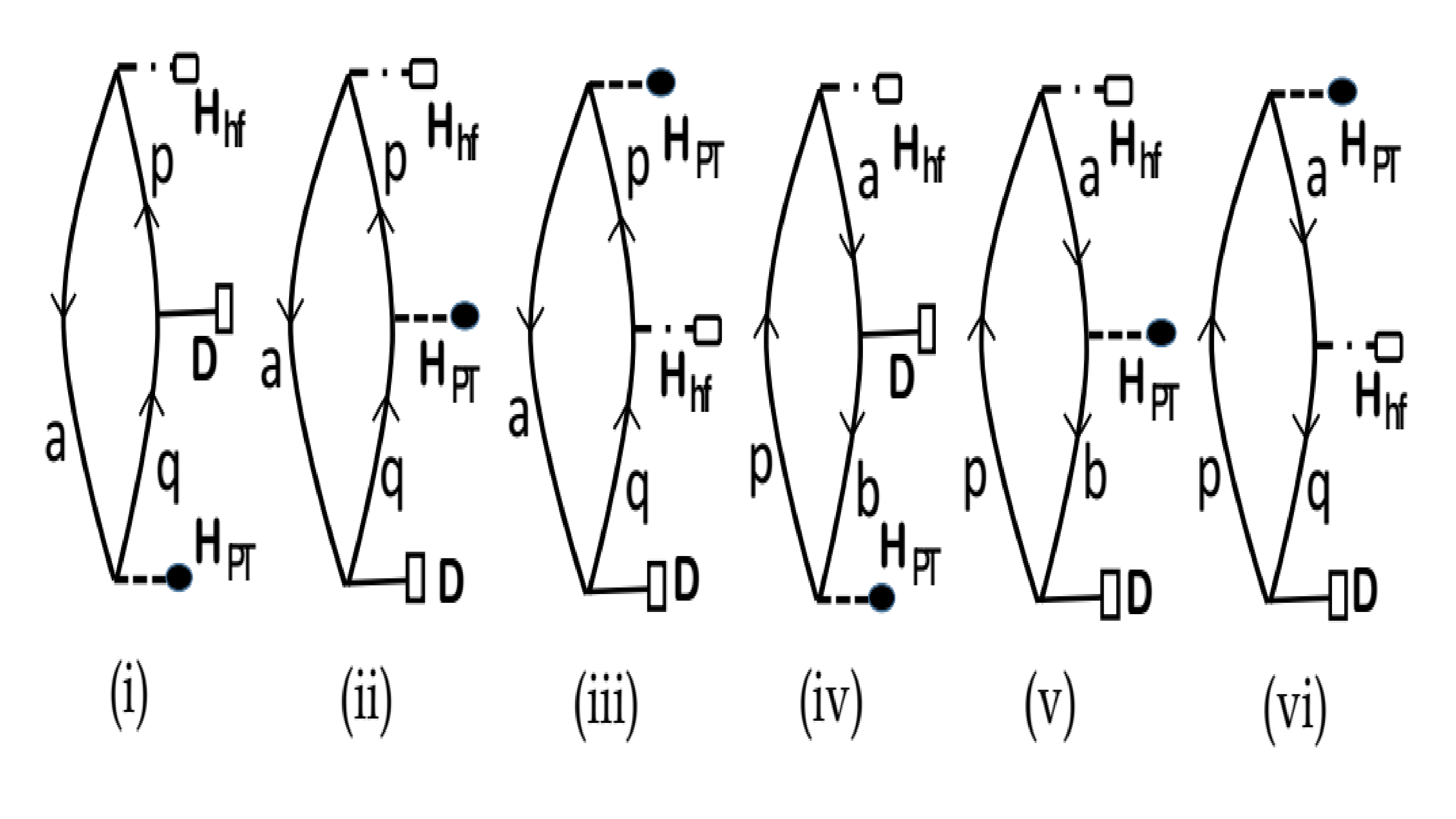}
\caption{Diagrammatic representation of different DHF contributions to the $d_\mathrm{a}^{3rd}$ values. In the figure, lines with upward arrows denote virtual orbitals and lines with downward arrows denote occupied orbitals. Operators $H_{hf}$, $H_{PT}$ and $D$ are shown by a singled dotted line with a rectangular box, a dotted line with black circle and a line with square respectively.}
\label{fig1}
\end{figure}

 \begin{table}[t]
 \caption{Contributions from different DHF diagrams to the $d_\mathrm{a}^{3rd}$ values using four representative basis functions. Values from $d_\mathrm{a}^e$ and $d_\mathrm{a}^{Sc}$ are given in $\times 10^{-4}$ e-cm and $\times 10^{-23}(C_\mathrm{S}/A)$ e-cm respectively. }
\begin{tabular}{l c rr r}
\hline \hline
   Fig. & Basis &  \multicolumn{2}{c}{$d_\mathrm{a}^e$ value} & $d_\mathrm{a}^{Sc}$ value\\
   \cline{3-4} \\ 
   No.         &   Set        &   This work & Ref. \cite{Martensson} &    \\
 \hline \\
 Fig. \ref{fig1}(i) &  I  &  $-0.878$ &   & $-0.054$ \\
                    &  II & $-0.874$    &        & $-0.054$  \\
                    &  III  & $-0.874$    &      & $-0.054$ \\
                    &   V   & $-0.872$ &  &  $-0.054$  \\
                    &  {\bf VIII} &  $-${\bf 0.872}  & 0.870  & $-0.054$ \\
 Fig. \ref{fig1}(ii) &  I  & 1.664  &  & 1.021 \\
                    &  II & 5.675    &        & 1.061\\
                    &  III  & 6.288 &   & 1.832  \\
                    &   V   & 6.338  &  & 1.833  \\
                    &  {\bf VIII} & {\bf 6.338}   & $-4.887$   & 1.833 \\
 Fig. \ref{fig1}(iii) &  I  & 3.109  &    & 0.200  \\
                    &  II &  7.170  &        & 1.203 \\
                    &  III  &  7.757     &   & 1.957 \\
                    &   V   & 7.948 &  & 1.959  \\
                    &  {\bf VIII} &  {\bf 7.948}  &  $-6.697$ & 1.959  \\
 Fig. \ref{fig1}(iv) &  I  & 0.890 &  &  0.055 \\
                    &  II & 0.892       &        &  0.055 \\
                    &  III  & 0.892 &  & 0.055 \\
                    &   V   & 0.893 &    & 0.055 \\
                    &  {\bf VIII} & {\bf 0.893}   & $-0.963$   &  0.055 \\
 Fig. \ref{fig1}(v) &  I  & $-2.870$  &   & $-0.172$\\
                    &  II & $-2.870$       &        & $-0.172$ \\
                    &  III  & $-2.870$  &      &  $-0.172$ \\
                    &   V   &  $-2.861$  &   &  $-0.171$ \\
                    &  {\bf VIII} & $-${\bf 2.861}   & 2.859   &  $-0.171$ \\
 Fig. \ref{fig1}(vi) &  I  & $-1.275$ &  & $-0.077$ \\
                    &  II & $-1.275$    &        & $-0.077$ \\
                    &  III  & $-1.275$  &   & $-0.077$ \\
                    &   V   & $-1.274$  &   &  $-0.077$ \\
                    &  {\bf VIII} & $-${\bf 1.274}  & 1.274   &  $-0.077$\\
\hline \hline
\end{tabular}
\label{tab2}
\end{table}

 \begin{table*}[t]
 \caption{Convergence of the RPA values of the estimated $\alpha_d$ and EDM enhancement factors from various P,T-odd interactions in $^{129}$Xe with different size of basis functions.}
\begin{tabular}{l cc cc cc cc}
\hline \hline
Set No. & Basis size & $\alpha_d$ & $d_\mathrm{a}^{Sm} \times 10^{-17}$ & $d_\mathrm{a}^T \times 10^{-20}$ & $d_\mathrm{a}^{Ps} \times 10^{-23}$ & $d_\mathrm{a}^B \times 10^{-4} $ & $d_\mathrm{a}^e \times 10^{-4} $ &  $d_\mathrm{a}^{Sc}\times 10^{-23}$ \\
&  & (a.u.) & (${S/(e \ \text{fm}^3)}$ e-cm) & ($\langle \sigma \rangle C_\mathrm{T}$ e-cm) & ($\langle \sigma \rangle C_\mathrm{P}$ e-cm) & e-cm & e-cm & ($(C_\mathrm{S}/A)$ e-cm) \\
 \hline \\
 I   &  $20s$, $20p$   & 6.753 & 0.481 & 0.723 & 2.088  & 1.036  &  0.541 &  0.052 \\
 II  &  $30s$, $30p$   & 6.753 & 0.482 & 0.723 & 2.088 & 1.031 & 13.582 & 3.234 \\ 
 III &  $35s$, $35p$   & 6.753 & 0.482 & 0.723 & 2.088 & 1.031 & 15.518 &  5.504 \\
 IV  &  $40s$, $40p$   & 6.753 & 0.482 & 0.723 & 2.088 & 1.031 & 15.519 &  5.509 \\
  V  &  $35s$, $35p$, $35d$  & 26.923 & 0.379 & 0.565 & 1.634 & 0.794 & 12.168 &  4.463 \\
 VI  &  $40s$, $40p$, $40d$  & 26.923 & 0.379 & 0.565 & 1.634 & 0.794 & 12.172 &  4.466 \\
VII &  $40s$, $40p$, $40d$, $40f$, $40g$  & 26.975 & 0.379 & 0.565 & 1.634 & 0.794 & 12.172 &  4.466 \\
 {\bf VIII}  & {\bf 35}$s$, {\bf 35}$p$, {\bf 15}$d$, {\bf 15}$f$, {\bf 15}$g$ & {\bf 26.975} & {\bf 0.378} & {\bf 0.564} & {\bf 1.631} & {\bf 0.795} & {\bf 12.168} &  {\bf 4.463} \\
 IX &  $20s$, $20p$, $20d$, $15f$, $15g$  & 26.975 & 0.378  &  0.564  & 1.631 & 0.795 & 0.441 &  0.051 \\
\hline \hline
\end{tabular}
\label{tab3}
\end{table*}

The S-Ps interaction Hamiltonian is given by
\begin{eqnarray}
H_{SPs} = \frac{i G_F C_S}{\sqrt{2}} A \sum_k \beta_k \gamma_k^5 \rho_\mathrm{nuc}(r) = \sum_k h_k^{SPs} ,
\end{eqnarray}
where 
%$G_F$ is the Fermi constant and 
$A$ is the atomic mass number of the considered atom. Matrix elements of its single particle operator $h^{SPs}$ is given by Eq. (\ref{eqsps}). Since the above interaction Hamiltonian is scalar in nature, it will contribute to EDM of a closed-shell atom through the hyperfine induced interaction. Thus, it can be evaluated using the expression
\begin{eqnarray}
d_\mathrm{a}^{Sc} &=& 2 {\lambda_1} {\lambda_2} \frac{ \langle \Psi_0^{(0,0)} | D  |\Psi_0^{(1,1) }\rangle + \langle \Psi_0^{(1,0)} | D  |\Psi_0^{(0,1) }\rangle }{\langle \Psi_0^{(0,0)} | \Psi_0^{(0,0)} \rangle}  .
\label{eqSc}
\end{eqnarray}

The Ps-S interaction interaction Hamiltonian is given by
\begin{eqnarray}
	H_{PsS} &=& -  \frac{ G_F C_P} {2 \sqrt{2} m_p c} \sum_{k} \gamma_0 \bm{\sigma}_\mathrm{nuc}   \nabla_k \rho_\mathrm{nuc}(r) \nonumber \\
      &=& \sum_k h_k^{PsS}(r) ,
	\label{eq:PSS}
\end{eqnarray}
where $m_p$ is the mass of a proton and $\bm{\sigma}_\mathrm{nuc}= \sum_n \langle \sigma_{n} \rangle + \sum_p \langle \sigma_{p} \rangle$ is the Pauli spin operator for the nucleus. Matrix element for its single particle operator $h^{PsS}$ is given by Eq. (\ref{eqpss}). Contribution to ${d_\mathrm{a}}$ from the above Hamiltonian is evaluated by
 \begin{equation}
 	d_\mathrm{a}^{Ps} \simeq 2{\lambda_2} \frac{ \langle \Psi_0^{(0,0)} | D  |\Psi_0^{(0,1)}\rangle }{\langle \Psi_0^{(0,0)} | \Psi_0^{(0,0)} \rangle} .
	\label{edmPs}
 \end{equation}

The T-Pt e-N interaction Hamiltonian for an atomic system is given by \cite{Sandars, Martensson1, V. V. Flambaum} 
\begin{eqnarray}
	H_\mathrm{int}^\mathrm{TPt} &=& i  \sqrt{2} G_F C_T \sum_{k} (  \bm{\sigma}_\mathrm{nuc} \cdot {\gamma}_k^0) \rho_\mathrm{nuc}(r) \nonumber \\
      &=& \sum_k h_k^{TPt}(r) ,
	\label{eq:TPT}
\end{eqnarray}
and the matrix element of its single particle operator is given by Eq. (\ref{eqtpt}).  Contribution to ${d_\mathrm{a}}$ from the above Hamiltonian is evaluated by
 \begin{equation}
 	d_\mathrm{a}^T \simeq 2{\lambda_2} \frac{ \langle \Psi_0^{(0,0)} | D  |\Psi_0^{(0,1)}\rangle }{\langle \Psi_0^{(0)} | \Psi_0^{(0)} \rangle} .
	\label{edmT}
 \end{equation}

We would like to mention here is that the $C_P$ coefficient can be deduced approximately from $C_T$ and vice versa using the relation
\begin{eqnarray}
C_P \approx 3.8 \times 10^3 \times \frac{A^{1/3}}{Z} C_T ,
\label{eqcp}
\end{eqnarray}
where $Z$ is the atomic number of the atom. However, reliability of this relation has not been verified yet. Thus, it would be necessary to infer both the coefficients separately to test the above relation.

\subsection{Methodology}

The RCC method is a non-perturbative theory to a many-body problem. Its notable characteristics are many folds compared to other contemporary many-body methods that are generally employed to carry out calculations of spectroscopic properties. Among them the main advantages of a RCC method is that its formulation satisfies size-consistent and size-extensivity properties, its ability to account for different types of correlation effects on equal footing (also cross correlations among them) and capturing more physical effects at the given level of approximation compared to other popular many-body methods \cite{helgaker,crawford,bartlett}. We employ this theory to estimate enhancement coefficients due to each of the P,T-odd interaction. Calculation of wave functions of an atomic system necessitates to obtain first a suitable mean-field wave function (reference state) including part of the electron correlation effects and treat the residual correlation effects as external perturbation. Thus, evaluating the second- and third-order EDM properties of an atomic system, as discussed in the previous section, means dealing with another source of perturbation along with the residual correlation effects. This makes it challenging to determine the intended properties using the RCC method.   

We consider the Dirac-Coulomb (DC) Hamiltonian to determine the unperturbed wave function $| \Psi_0^{(0,0)} \rangle$ due to the dominant electromagnetic interactions, given by
\begin{equation}
	H_0 = \sum_{i}^{N_e} [c\bm{\alpha} \cdot \bm{p}_i + c^2 \bm{\beta} + V_\mathrm{nucl} (r_i) ] +\cfrac{1}{2} \sum_{i,j} \cfrac{1}{r_{ij}},
\end{equation}
where $N_e$ is the number of electrons, $\bm{\alpha}$ is the Dirac matrix, $V_\mathrm{nucl} (r_i)$ is the nuclear potential, and $r_{ij}$ is the distance between $i^{th}$ and $j^{th}$ electrons. In the above expression, we have used atomic units (a.u) in which $\hbar=1$ and mass of electron $m_e=1$.

In the RCC theory framework, we can express $|\Psi_0^{(0,0)} \rangle $ due to $H_0$ as
\begin{equation}
	|\Psi_0^{(0,0)} \rangle = e^{T^{(0,0)}} | \Phi_0 \rangle,
	\label{eqrcc}
\end{equation}
where $|\Phi_0\rangle$ is the mean-field wave function obtained using the Dirac-Hartree-Fock (DHF) method and the cluster operator $T^{(0,0)}$ is defined as
 \begin{equation}
 	T^{(0,0)} = \sum_{I=1}^{N_e} T_I^{(0,0)} = \sum_{I=1}^{N_e} t_I^{(0,0)}  C_I^{+},
 \end{equation}
where $I$ represents the number of particle-hole pairs, $t_I^{(0,0)}$ is the unperturbed excitation amplitude, and $C_I^+$ is the $I$ pair of creation and annihilation operators denoting level of excitations. In our work, we have considered singles and doubles approximation in the RCC theory (RCCSD method) by restricting $I$ up to one-particle--one-hole and two-particle--two-hole excitations; i.e. $T^{(0,0)}=T_1^{(0,0)} + T_2^{(0,0)}$.  The general $T^{(0)}$ amplitude solving equations in the RCC theory is given by
  \begin{equation}
  	\langle \Phi_0 |C_I^{-}\overline{H}_0  |\Phi_0 \rangle = 0,
   \label{eqt0}
  \end{equation}
 where $C_I^{-}$ are the adjoint of  $C_I^{+}$ (referred to de-excitation) and $\overline{H}_0 = e^{-T^{(0,0)}} H_0 e^{T^{(0,0)}} = (H_0 e^{T^{(0,0)}})_l$ with subscript $l$ denoting for the linked terms (here onwards we shall follow the notation $\overline{O} =  (O e^{T^{(0,0)}})_l$ throughout the paper). Since $H_0$ has only one-body and two-body terms, $\overline{H}_0$ can have finite number of terms. In the RCCSD method approximation, we can have two set of equations for $T_1^{(0,0)}$ and $T_2^{(0,0)}$ as
\begin{eqnarray}
  	\langle \Phi_0 |C_1^{-} (H_0 T_1^{(0,0)})_l |\Phi_0 \rangle = - \langle \Phi_0 |C_1^{-} H_0 + (H_0 T_2^{(0,0)})_l |\Phi_0 \rangle  \nonumber \\
     - \langle \Phi_0 |C_1^{-}  \left [ H_0 \sum_{n,m} \frac{T_1^{(0,0)n} T_2^{(0,0)m}}{n! m!}  \right ]_l |\Phi_0 \rangle    \ \ \ \ 
\end{eqnarray}
and
\begin{eqnarray}
  	\langle \Phi_0 |C_2^{-} (H_0 T_2^{(0,0)})_l |\Phi_0 \rangle = - \langle \Phi_0 |C_2^{-} H_0 + (H_0 T_1^{(0,0)})_l |\Phi_0 \rangle  \nonumber \\
     - \langle \Phi_0 |C_2^{-}  \left [ H_0 \sum_{n,m} \frac{T_1^{(0,0)n} T_2^{(0,0)m}}{n! m!}  \right ]_l |\Phi_0 \rangle ,   \ \ \ \ 
\end{eqnarray} 
 where $n,m \ge 1$ denoting all possible non-linear terms. The above equations are solved using the Jacobi iterative procedure.

Now considering external perturbations due to $M1_{hf}$ and $H_{PT}$, we can express the total Hamiltonian as 
\begin{equation}
	H = H_0 + \lambda_1 M1_{hf} + \lambda_2 H_{PT} .
\end{equation}
In the RCC theory framework, we can express $|\Psi_0 \rangle$ of $H$ in the form similar to the unperturbed wave function as
\begin{equation}
	|\Psi_0 \rangle = e^T | \Phi_0 \rangle .
	\label{eqrcc1}
\end{equation}
In order to obtain the perturbed wave functions from this expression, we can express
\begin{equation}
T \simeq T^{(0,0)} + \lambda_1 T^{(1,0)} + \lambda_2 T^{(0,1)} + \lambda_1 \lambda_2 T^{(1,1)} ,
\end{equation}
where superscript notations are as per Eq. (\ref{eqdpt}). This follows 
\begin{eqnarray}
  &&  |\Psi_0^{(1,0)} \rangle = e^{T^{(0,0)}} T^{(1,0)} | \Phi_0 \rangle , \nonumber \\
  &&  |\Psi_0^{(0,1)} \rangle = e^{T^{(0,0)}} T^{(0,1)} | \Phi_0 \rangle \nonumber \\
    \text{and} && \nonumber \\
  && |\Psi_0^{(1,1)} \rangle = e^{T^{(0,0)}} \left ( T^{(1,1)} + T^{(1,0)} T^{(0,1)}\right ) | \Phi_0 \rangle . \ \ \
\end{eqnarray}

The amplitudes of the perturbed RCC operators can be obtained as
  \begin{eqnarray}
\langle \Phi_0 |C_I^{-} \left [ \overline{H}_0 T^{(1,0)} + \overline{M1}_{hf} \right ] | \Phi_0\rangle &=& 0 , \nonumber \\
\langle \Phi_0 |C_I^{-} \left [ \overline{H}_0 T^{(0,1)} + \overline{H}_{PT} \right ] | \Phi_0\rangle &=& 0  \nonumber 
  \end{eqnarray}
  and 
\begin{eqnarray}
 \langle \Phi_0 |C_I^{-} \left [ \overline{H}_0 T^{(1,1)} +  \overline{H}_0 T^{(1,0)} T^{(0,1)} \right. \nonumber \\ 
 \left. + \overline{M1}_{hf} T^{(0,1)} + \overline{H}_{PT} T^{(1,0)} \right ] | \Phi_0\rangle &=& 0 . 
\end{eqnarray}
It should be noted that the first two-equations are independent from each other and are solved separately after obtaining $T^{(0,0)}$ amplitudes. These two equations are of similar form with Eq. (\ref{eqt0}), so they are also solved using the Jacobi iterative procedure. Once amplitudes of the $T^{(0,0)}$,  $T^{(1,0)}$ and $T^{(0,1)}$ operators are known then amplitudes of the $T^{(1,1)}$ operator are obtained by solving the last equation in the same Jacobi iterative approach. Since $\overline{O}$ contains many non-linear terms among which $H_0$ also contains two-body terms, we use intermediate computational schemes to solve the amplitude determining equation for $T^{(1,1)}$. We divide $\overline{H}_0$ into effective one-body and two-body terms like the bare Hamiltonian $H_0$, and store them to use further for solving all three equations. This reduces a lot of computational time to obtain the perturbed RCC operator amplitudes. Due to limitation in memory of the available computational facility, it is not possible to store additional effective two-body terms that could arise from $\overline{M1}_{hf}$ and $\overline{H}_{PT}$. Since both $M1_{hf}$ and $H_{PT}$ are one-body operators, less number of two-body terms will arise from $\overline{M1}_{hf}$ and $\overline{H}_{PT}$ compared to $\overline{H}_0$. Thus, their effective one-body diagrams are only computed and stored for further use in the above equations, while their effective two-body terms are computed directly. In the last equation, we compute effective one-body terms of $ \overline{H}_0 T^{(1,0)} + \overline{M1}_{hf}$ together then multiplied by $T^{(0,1)}$ to compute the $\overline{H}_0 T^{(1,0)} T^{(0,1)}$ and $\overline{M1}_{hf} T^{(0,1)}$ terms economically. In the RCCSD method approximation, we write 
\begin{eqnarray}
T^{(1,0)}&=&T_1^{(1,0)} + T_2^{(1,0)} , \nonumber \\
T^{(0,1)}&=&T_1^{(0,1)} + T_2^{(0,1)} \nonumber 
\end{eqnarray}
and
\begin{eqnarray}
T^{(1,1)}&=&T_1^{(1,1)} + T_2^{(1,1)} .
\end{eqnarray}

With the knowledge of $T^{(1,0)}$, $T^{(0,1)}$ and $T^{(1,1)}$ amplitudes, we can evaluate the second-order EDM enhancement factors as 
 \begin{eqnarray}
  \frac{d_\mathrm{a}^{2nd}}{\lambda_2} &\simeq& 2 \frac{\langle \Phi_0 |{e^{T^{(0,0)}}}^{\dagger} D e^{T^{(0,0)}}T^{(0,1)} | \Phi_0 \rangle} {\langle \Phi_0 |{e^{T^{(0,0)}}}^{\dagger} e^{T^{(0,0)}}| \Phi_0 \rangle} \nonumber \\
	 &\simeq & 2 \langle \Phi_0 |\widetilde{D} T^{(0,1)} | \Phi_0 \rangle_l ,
	\label{eqedm}
 \end{eqnarray}
where $\widetilde{D}= {e^{T^{(0,0)}}}^{\dagger} D e^{T^{(0,0)}}$. As can be seen, the normalization of wave function has been cancelled with the unlinked terms of $\widetilde{D}$ in the above expression leaving out only the linked terms for the final evaluation. This argument can be followed from the discussions given in Refs. \cite{Yashpal, Bijaya} and the this is further verified using the biorthogonal condition \cite{bijaya2,sakurai}. Proceeding in the similar manner, the third-order EDM enhancement factors can be evaluated using the expression
  \begin{eqnarray}
  \frac{d_\mathrm{a}^{3rd}}{\lambda_1 \lambda_2}  & \simeq& 2 \langle \Phi_0 | \widetilde{D} T^{(1,1)} + {T^{(1,0)}}^{\dagger} \widetilde{D} T^{(0,1)} | \Phi_0 \rangle_l .
	\label{eqedm1}
 \end{eqnarray}
 We adopt an iterative procedure to evaluate contributions from $\widetilde{D}$ self-consistently. Once $\widetilde{D}$ is computed and stored, each term is reduced to a terminated expression in both Eqs. (\ref{eqedm}) and (\ref{eqedm1}) in the RCCSD method approximation to obtain the final result.

\begin{table*}
\caption{Contributions to $\alpha_d$ and $d_\mathrm{a}^{2nd}$ enhancement factors from various P,T-odd interactions in $^{129}$Xe through individual terms of the RCCSD method. The terms that are not shown explicitly their contributions are given together under ``Others". Estimated contributions from the Breit and QED interactions are given in the bottom of the table.}
\begin{tabular}{l cc cc cc}
\hline \hline
 RCC terms &  $\alpha_d$ & $d_\mathrm{a}^{Sm} \times 10^{-17}$ & $d_\mathrm{a}^T \times 10^{-20}$ & $d_\mathrm{a}^{Ps} \times 10^{-23}$ & $d_\mathrm{a}^B \times 10^{-4} $  \\
 & (a.u.) & (${S/(e \ \text{fm}^3)}$ e-cm) & ($\langle \sigma \rangle C_\mathrm{T}$ e-cm) & ($\langle \sigma \rangle C_\mathrm{P}$ e-cm) & e-cm  \\
 \hline \\
$ D T_1^{(0,1)} + \text{h.c.} $   &  29.980 & 0.318 &	0.510 & 1.471  & 0.722 \\ 
$ {T_1^{(0,0)}}^{\dagger} D T_1^{(0,1)} + \text{h.c.}$  & $-0.345$ &  0.003 & 	0.004 & 0.017 & 0.007 \\ 
$ {T_2^{(0,0)}}^{\dagger} D T_1^{(0,1)}+ \text{h.c.} $  & $-3.308$ &  0.011 & 0.017 &  0.049 & 0.034\\ 
$ {T_1^{(0,0)}}^{\dagger} D T_2^{(0,1)} +\text{h.c.} $  & 0.074  &   $\sim 0.0$ & $\sim 0.0$ & $-0.001$ & $-0.001$\\ 
${T_2^{(0,0)}}^{\dagger} D T_2^{(0,1)} + \text{h.c.} $	&  1.072  	&  $\sim 0.0$ & $\sim 0.0$ & $-0.001$ & $-0.003$\\ 
Others  & 0.042  &  0.013 & $-0.009$  & $-0.031$ & $-0.014$ & \\
\hline  \\
Breit & 0.051 & $-0.002$  & $-0.001$  & $-0.003$ & 0.003 \\
QED   & $-0.015$ & $-0.006$  & $-0.011$ & $-0.059$  & $-0.032$ \\
\hline \hline
\end{tabular}
\label{tab4}
\end{table*}

\section{Results and discussion}

Before presenting the results from various P,T-odd interaction sources to EDM of $^{129}$Xe, it would be important to validate the calculations. There are two aspects to be looked into in such intent -- completeness of basis functions used in the generation of atomic orbitals and reproducing some known quantities (i.e. comparing between the calculated and experimental results) using the determined wave functions. It is very tactful business to deal with basis functions in the calculations of atomic properties as it is not possible to obtain a complete set of basis functions to estimate a property of our interest. In the consideration of finite-size basis functions, they are chosen keeping in view of sensitivity of a given property at the shorter or longer radial distances. Matrix elements of the $D$ operator are more sensitive to the wave functions at longer distances. However, the P,T-odd interactions of our interest are originating from the nucleus. The $s$ and $p_{1/2}$ orbital wave functions having larger overlap with the nucleus are supposed to be contributing predominantly to the matrix elements of $H_{PT}$. It may not be necessary to use sufficient number of orbitals from higher orbital angular momentum; $l>1$. Again, energy denominators can also play crucial roles in deciding important contributing high-lying orbitals to the perturbative quantities. Thus, it is expected that contributions from the $ns$ and $np_{1/2}$ orbitals to EDM with principal quantum number $n>20$ may not be large. This argument may be valid in the determination of the $d_\mathrm{a}^{2nd}$ values, but one has to be careful with such presumption in the evaluation of the $d_\mathrm{a}^{3rd}$ contributions. This is because the third-order contributions to EDM of $^{129}$Xe can be enhanced by the $\langle ns | M1_{hf} | ms \rangle$ and $\langle np_{1/2} | M1_{hf} | mp_{1/2} \rangle$ matrix elements with continuum orbitals lying beyond $n,m>20$ due to the fact that these orbitals have large overlap within the nuclear region, and energy differences between the associated $ns$ and $np_{1/2}$ orbitals do not appear in the denominator of the terms involving the $\langle ns | M1_{hf} | ms \rangle$ and $\langle np_{1/2} | M1_{hf} | mp_{1/2} \rangle$ matrix elements. It is possible to verify enhancement to the EDM contributions from these high-lying orbitals using the DHF method or using an all-order method like random phase approximation (RPA), as these methods do not require much computational resources. The point about determining some quantities and comparing them with their experimental values, it would be desirable to search for properties having similarities with the EDM calculations. However, Evaluation of EDM involves matrix elements of $D$, matrix elements of $H_{PT}$ (via $|\Psi_0^{(0,1)} \rangle$ and $|\Psi_0^{(1,1)} \rangle$) and excitation energies (appearing in the denominator of the amplitude coefficients of the perturbed wave function) and there is no such measurable property of $^{129}$Xe known which has striking similarity with the calculation of its EDM. In the open-shell EDM studies, one evaluates hyperfine structure constants and electric dipole polarizabilities ($\alpha_d$) obtained using the calculated wave functions to compare them with their available experimental values for testing accuracy of the atomic wave functions in the nuclear and asymptotic regions, respectively. Since the ground state of $^{129}$Xe does not have hyperfine splitting, we only determine its $\alpha_d$ and compare it with the experimental value. The same has also been done earlier while calculating contributions from P,T-odd interactions to atomic EDM of $^{129}$Xe \cite{Dzuba,Yashpal,Sakurai,Fleig}. 

It is well known in the literature that Gaussian type of orbitals (GTOs) form a good set of basis functions that can describe wave functions near the nuclear region very well \cite{Boys,Mohanty,Dyall}. We have also used Fermi nuclear charge distribution \cite{Estevez} to define $\rho_N(r)$ and nuclear potential. We have used 40 GTOs using even tempering condition, as described in \cite{Schmidt}, for each orbital belonging to $l$ values up to 4 (i.e. $g$-symmetry) in the present calculations. There are two reasons for not considering orbitals from the higher momentum values. First, these omitted orbitals do not contribute up to the desired precision to the EDM of $^{129}$Xe. Second, evaluation of $d_\mathrm{a}^{3rd}$ demands for inclusion of higher $s$ and $p$ continuum orbitals to obtain reliable results for EDM. So inclusion of higher angular momentum orbitals to account for electron correlation effects in the RCCSD method would be a challenge with the available computational facilities, especially orbitals from $l>4$ that do not contribute significantly to the matrix elements of $H_{PT}$. We also demonstrate in this work that how a set of basis function that would be sufficient to provide accurate value of $\alpha_d$ is not sufficient enough to estimate $d_\mathrm{a}^{3rd}$ contributions correctly. In view of the aforementioned discussions, it would be necessary to investigate convergence of $d_\mathrm{a}^{3rd}$ contributions to EDM by considering as many $ns$ and $np_{1/2}$ orbitals as possible in the calculations. 

In Table \ref{tab0}, we summarize the calculated $\alpha_d$, $d_\mathrm{a}^{2nd}$ and $d_\mathrm{a}^{3rd}$ values of $^{129}$Xe from the DHF, RPA and RCCSD methods. The reason for giving results from RPA is, the previous calculations were mostly reported results using this approach. Again, differences between the DHF and RPA results will indicate the roles of core-polarization contributions while differences in the RPA and RCCSD results would exhibit the roles of non-core-polarization contributions in the determination of the investigated quantities. It can be seen from the table that differences between the DHF, RPA and RCCSD values are not so significant though non-negligible in all the evaluated properties. It means that correlation effects in this atom is not very strong. It can also be noticed that the $\alpha_d$ value increases from the DHF method to RPA, then from RPA to the RCCSD method. However, the $d_\mathrm{a}^{2nd}$ values show different trends -- these values increase from the DHF method to RPA then they decrease slightly in the RCCSD method. Since the RCCSD method implicitly contains all the RPA effects \cite{Yashpal}, it implies that the non-RPA effects arising through the RCCSD method behave differently in $\alpha_d$ and $d_\mathrm{a}^{2nd}$. The $d_\mathrm{a}^{3rd}$ values also show similar trends; i.e. first they increase from the DHF method to RPA then decrease slightly in the RCCSD method. However, correlation effects are relatively smaller in magnitude for the $d_\mathrm{a}^{3rd}$ values compared to the $d_\mathrm{a}^{2nd}$ values. Therefore, it is very important that the DHF values for $d_\mathrm{a}^{3rd}$ are determined reliably in order to estimate their final values more accurately using the RCCSD method. We also give our final values along with their possible uncertainties from the neglected contributions. These final results are estimated by including contributions from the Breit and lower-order QED interactions to the RCCSD values. These values are compared with the previous calculations reported in Refs. \cite{Martensson,Yashpal,sakurai,Fleig,Dzuba,Flambaum}. The calculated $\alpha_d$ values from the same methods, that are employed to obtain EDM results, are also compared with the experimental result \cite{hohm} in the above table. It shows that our calculated value $\alpha_d$ agrees well with the experimental result. They also match with our previous calculations \cite{Yashpal,sakurai}, where smaller size basis functions were used and contributions from the Breit and QED effects were neglected. However, our $\alpha_d$ value differs substantially from the value reported in Ref. \cite{Fleig} using the configuration interaction (CI) method. In fact, the CI value is found to be smaller than our DHF and RPA results. From the comparison of EDM results, we find our RPA values for $d_\mathrm{a}^{Sm}$, $d_\mathrm{a}^T$ and $d_\mathrm{a}^{Ps}$ match with the RPA values listed in Ref. \cite{Dzuba}. However, we find our RPA value for $d_\mathrm{a}^{B}$ differs from Ref. \cite{Dzuba} while it is almost in agreement with the RPA value given in Ref. \cite{Martensson}. A careful analysis of this result suggests that calculation of $d_\mathrm{a}^{B}$ is very sensitive to the choices of root mean square radius $R$  and radial integral limits in the evaluation of the single matrix elements of $h_k^B$ as demonstrated explicitly later. Our RCCSD values for all these quantities agree with the RCCSD results and calculations using the normal relativistic coupled-cluster theory reported in Refs. \cite{Yashpal,sakurai}. 

After discussing the second-order perturbative properties, we now move on to discussing the $d_\mathrm{a}^e$ and $d_\mathrm{a}^{Sc}$ values. Unlike the earlier discussed properties, we find our third-order properties differ significantly from the previously reported values. The reported $d_\mathrm{a}^e$ value in Ref. \cite{Martensson} was performed at the RPA level, while it was obtained analytically in Ref. \cite{Flambaum}. The $d_\mathrm{a}^{Sc}$ value of Ref. \cite{Fleig} was estimated using the CI method. In the case of $d_\mathrm{a}^e$, we observe a sign difference between our result and that are reported in Refs. \cite{Martensson,Fleig}. On other hand, the signs of our calculated $d_\mathrm{a}^{Sc}$ value agrees with the result of Ref. \cite{Fleig}. Since there is an analytical relationship between the S-Ps and electron EDM P,T-odd interaction Hamiltonians, signs of both the contributions are anticipated to be the same. From this analysis, we assume that sign of our estimated value is $d_\mathrm{a}^e$ is alright. Now looking into large differences in the magnitudes for these $d_\mathrm{a}^{3rd}$ contributions, we find that they are owing to different basis functions used in the calculations. This can also be corroborated from the fact that the correlation effects arising through the RCCSD method to the $d_\mathrm{a}^{3rd}$ contributions are not so much large, thus the main differences in the results come from the DHF values. The magnitudes of the $d_\mathrm{a}^e$ value among various calculations almost agree but there is an order magnitude difference for $d_\mathrm{a}^{Sc}$. The authors have analyzed roles of basis functions in the determination of $\alpha_d$, $d_\mathrm{a}^T$ and $d_\mathrm{a}^{Sc}$ in Ref. \cite{Fleig}. They have noticed large fluctuations in the results, and their final $\alpha_d$ value (i.e. 25.58 a.u) differs significantly from the experiment. Also, they have made a small virtual cut-off to manage the calculations with limited computational resources as the CI method can demand huge RAM in the computers for direct diagonalization of a bigger CI matrix. We demonstrate below using both the DHF and RPA methods how such cut-off for the virtual orbitals do not affect significantly to the determination of the $d_\mathrm{a}^{2nd}$ values, but they are very sensitive to the evaluation of $d_\mathrm{a}^{3rd}$ values. 

We present the DHF values for $\alpha_d$, $d_\mathrm{a}^{2nd}$ and $d_\mathrm{a}^{3rd}$ of $^{129}$Xe in Table \ref{tab1} from a different set of single particle orbitals. Since $s$, $p_{1/2}$ and $p_{3/2}$ orbitals are the dominantly contributing orbitals, we consider these orbitals first and gradually include orbitals with higher orbital angular momentum values till the $g$-symmetries to show that their roles in the determination of above quantities. At this stage it is important to note that some of the orbitals from higher angular momentum orbitals may not contribute through the DHF method but they can contribute via the electron correlation effects to the above quantities. Thus, if the correlation effects are significant only then one needs to worry about the contributions from the higher angular momentum (belonging to $l>4$) to the investigated properties. Anyway, we shall present variation of correlation effects through the RPA method considering a few typical set of orbitals later to show how inclusion of orbitals from the higher angular momentum can modify the results. In Table \ref{tab1}, we start presenting results considering 20$s$, 20$p_{1/2}$ and 20p$_{3/2}$ orbitals (set I). This is a reasonable size basis functions when only $s$ and $p$ orbitals make contributions to a property. Results reported from this set of basis functions are already close to the DHF values for all the $d_\mathrm{a}^{2nd}$ values, whereas there is a large difference for the $\alpha_d$ value from the final value of the DHF method as quoted in Table \ref{tab1}. We also see quite significant differences for the $d_\mathrm{a}^{3rd}$ values at the DHF method compared to what are listed in Table \ref{tab1}. This shows that contributions from other orbitals are also substantial to the evaluation of the $\alpha_d$ and $d_\mathrm{a}^{3rd}$ values, but their contributions are small for $d_\mathrm{a}^{2nd}$. To learn how the higher $ns$ and $np$ continuum orbitals, or orbitals with the higher orbital angular momentum can affect the results, we consider two more set of basis functions next including the 35$s$ and 35$p$ orbitals (set II) then increase up to 40$s$ and 40$p$ orbitals (set III). It shows that none of the $d_\mathrm{a}^{2nd}$ values as well as $\alpha_d$ make much change with the inclusion of more number of $ns$ and $np$ orbitals, but the $d_\mathrm{a}^{3rd}$ values change by one order with the inclusion of 35$s$ and 35$p$ orbitals and these values get saturated after that. This strongly advocates for the fact that roles of continuum orbitals beyond $n>20$ are very crucial for accurate estimation of the $d_\mathrm{a}^{3rd}$ values. We proceed further by adding orbitals from the higher angular momentum. We consider 35$d$ orbitals first along with 35$s$ and 35$p$ orbitals (set IV) then 40$d$ orbitals along with 40$s$ and 40$p$ orbitals (set V). The DHF values in both the cases seem to be almost same for all these quantities. Compared with the previous set of orbitals, we find none of the $d_\mathrm{a}^{2nd}$ and $d_\mathrm{a}^{3rd}$ values are changed except the $\alpha_d$ value. This asserts our earlier statement about how EDM results are sensitive to only the higher $ns$ and $np$ orbitals but contributions from other orbitals to EDM are negligibly small. Nonetheless, orbitals from the $g$ symmetry do not contribute to the DHF method as there are no occupied orbitals in the $f$ shell of $^{129}$Xe while virtual $f$ orbitals can contribute due to presence of the occupied $d$ orbitals. Their contributions to EDM are negligible while a small contribution from these orbitals is noticed to the determination of $\alpha_d$. 

\begin{table}[t]
\caption{Contributions to the $d_\mathrm{a}^{3rd}$ enhancement factors from the electron EDM and S-PS interactions in $^{129}$Xe through individual terms of the RCCSD method. The terms that are not shown explicitly their contributions are given together as ``Others". The Breit and QED interaction contributions are given in the end of the table.}
\begin{tabular}{l cc}
\hline \hline
 RCC terms  &  $d_\mathrm{a}^e \times 10^{-4} $ &  $d_\mathrm{a}^{Sc}\times 10^{-23}$ \\
&   e-cm & ($(C_\mathrm{S}/A)$ e-cm) \\
 \hline \\
$ D T_1^{(1,1)} + \text{h.c.} $   &  10.922  & 3.953 \\ 
$ {T_1^{(0,1)}}^{\dagger} D T_1^{(1,0)} + \text{h.c.}$  &  $-0.076$	 & $-0.004$	\\ 
$ {T_2^{(0,1)}}^{\dagger} D T_1^{(1,0)}+ \text{h.c.} $  &  $-0.045$  &  $-0.003$ \\ 
$ {T_1^{(0,1)}}^{\dagger} D T_2^{(1,0)} +\text{h.c.} $  &  0.0  	&  0.0 \\   
${T_2^{(0,0)}}^{\dagger} D T_2^{(1,1)} + \text{h.c.} $	&  $-0.018$ &   $-0.002$\\ 
${T_2^{(0,1)}}^{\dagger} D T_2^{(1,0)} + \text{h.c.} $	&  $-0.020$  	&  $-0.001$ \\
Others & 0.428 & 0.088 \\
\hline  \\
Breit & $-0.037$ &  $-0.008$  \\
QED   & $-0.417$ & $-0.118$ \\
\hline  \hline   
\end{tabular}
\label{tab5}
\end{table}	

In the present work, we have used Fermi type nuclear charge distribution, given by
\begin{equation}
    \rho(r)=\frac{\rho_0}{1+e^{(r-b)/a}},
    \label{2pF}
\end{equation}
where $\rho_0$ is a normalization constant, $b$ is the half-charge radius and $a=2.3/4ln(3)$ is related to the skin thickness. The relation between $R$, $b$ and $a$ are given by
\begin{eqnarray} 
R = \sqrt{ \frac{3}{5} b^2 + \frac{7}{5} a^2 \pi^2 }.
\end{eqnarray}
In Table \ref{tab02}, we show how the DHF value for $d_\mathrm{a}^{B}$ changes with $R$ (by varying $b$ value) and cut-off in the radial integration of the wave functions with basis set VIII. As can be seen from the table, for a small radial integral cut-off the results show opposite signs than for the larger cut-offs. The value increases till 200 a.u. then slightly decrease at the very large cut-off value. Beyond 500 a.u., we do not see any further changes in the results. Again, we see significant variation in the results with $b$ values. In our calculation, we use $b=5.655$ fm at which it satisfies the empirical relation
\begin{eqnarray}
R = 0.836 A^{1/3} + 0.570 \ \text{fm} ,
\end{eqnarray}
where $A$ is the atomic mass of $^{129}$Xe. Thus, one of the reasons for the difference in the $d_\mathrm{a}^{B}$ value between the present work and that are reported in \cite{Flambaum,Martensson} could be due to choices of different nuclear charge radius and cut-off in the radial integration of the matrix elements.

We also verify how the hyperfine-induced results differ without and with considering magnetization distribution (${\cal M}(r)$) within the nucleus. In this case too, we use Fermi type distribution as
\begin{equation}
    {\cal M}(r)=\frac{1}{1+e^{(r-b)/a}} .
    \label{Mdf}
\end{equation}
The DHF values for $d_\mathrm{a}^e$ and $d_\mathrm{a}^{Sc}$ without and after multiplying the above factor with the $M1_{hf}$ operator are given in Table \ref{tab03}. As can be seen from the table, there are significant reduction in the magnitudes of the above quantities when magnetization distribution is taken into account within the nucleus. Our final results reported in Table \ref{tab1} include these effects. 

In order to analyze how the high-lying orbitals enhance the $d_\mathrm{a}^{3rd}$ contributions in the DHF method, we take the help of Goldstone diagrams as have been described in Ref. \cite{Martensson}. In Fig. \ref{fig1}, we show these Goldstone diagrams representing six terms of the DHF method that contribute to $d_\mathrm{a}^{3rd}$. We present contributions from these diagrams in Table \ref{tab2} using four representative set of basis functions that are denoted as sets I, II, III, V and VIII in Table \ref{tab1}. We have also compared our results diagram-wise from the bigger basis (set VIII) with the results from Ref. \cite{Martensson}. As can be seen from the table, result from set I that gives very small DHF values to $d_\mathrm{a}^{3rd}$ produces reasonable contributions through via  Figs. \ref{fig1}(i) and (iv), (v) and (vi). In all these cases, matrix elements of $H_{PT}$ and $M1_{hf}$ are involved with at least one core orbital. The remaining two diagrams involve matrix elements of $H_{PT}$ and $M1_{hf}$ between virtual orbitals whose energy denominators do not appear in the evaluation of the DHF value. This ascertains our initial discussion about why high-lying virtual orbitals enhance the $d_\mathrm{a}^{3rd}$ contributions. Compared to results from Ref. \cite{Martensson}, we find our results from Figs. \ref{fig1}(i), (v) and (vi) match quite well (only the magnitude, but sign differs as was mentioned earlier) while they differ for the other diagrams. We also find trends in the results from different DHF diagrams are different for $d_\mathrm{a}^e$ and $d_\mathrm{a}^{Sc}$. This is clearly evident from the contributions of Figs. \ref{fig1}(ii) and (iii), where basis sets I and II give small values for both the quantities. With basis set VIII, contributions to the $d_\mathrm{a}^e$ value becomes almost triple times larges while it only increases marginally for $d_\mathrm{a}^{Sc}$. Thus, it is evident from these discussions that choice of basis functions for the hyperfine-induced contributions to atomic EDMs seem to be very crucial.

As stated earlier, correlation effects between the $d$, $f$ and $g$ orbitals through the DHF potential is absent for the calculations above quantities. However, their correlation effects through the residual Coulomb interaction may affect the results through the RPA and RCCSD methods. To verify this fact, we make similar analysis in the trends of results by performing calculations with different set of basis functions using the RPA. These results are listed in Table \ref{tab3} from which it can be seen that the all-order method also show similar trends in the results as in the DHF method. From this exercise it follows that orbitals with higher angular momentum do not contribute significantly to the $d_\mathrm{a}^{2nd}$ and $d_\mathrm{a}^{3rd}$ contributions and consideration of high-lying $ns$ and $np$ orbitals with $n>20$ is essential for accurate estimate of the $d_\mathrm{a}^{3rd}$ contributions.

In Table \ref{tab4}, we present contributions from individual terms of the RCCSD method to the estimations of $\alpha_d$ and $d_\mathrm{a}^{2nd}$ values from different $H_{PT}$. we find that $D T_1^{(0,1)}$ and its hermitian conjugate (h.c.) gives almost all the contributions to the above quantities. The next dominant contributions arise through $ {T_2^{(0,0)}}^{\dagger} D T_1^{(0,1)}$ and its h.c.. Contributions from the higher-order non-linear terms, quoted as ``Others", are non-negligible. In the end of table, we have also listed contributions arising through the Breit and lower-order QED interactions. They show that Breit interaction contributes more to $\alpha_d$ than QED, while it is other way around for $d_\mathrm{a}^{2nd}$.

We also present contributions from the individual terms of the RCCSD method to the estimations of the $d_\mathrm{a}^{3rd}$ values in Table \ref{tab5}. In this case, the $D T_1^{(1,1)}+\text{h.c.}$ terms contribute mostly to both $d_\mathrm{a}^e$ and $d_\mathrm{a}^{Sc}$, and the next leading order contributions arise from $ {T_1^{(0,1)}}^{\dagger} D T_1^{(1,0)} + \text{h.c.}$. There are non-negligible contributions from $ {T_2^{(0,1)}}^{\dagger} D T_1^{(1,0)}+ \text{h.c.} $, ${T_2^{(0,0)}}^{\dagger} D T_2^{(1,1)} + \text{h.c.} $ and ${T_2^{(0,1)}}^{\dagger} D T_2^{(1,0)} + \text{h.c.} $. The rest of contributions, given as ``Others", are also quite significant.	In the bottom of the table, we quote contributions from both the Breit and QED interactions. Contributions arising through the QED interactions seem to be relatively large.
	
The latest reported experimental result for the EDM of $^{129}$Xe is 
 \cite{Allmendinger:2019jrk,Sachdeva:2019blc}
\begin{equation}
|d_{\rm Xe} | < 1.4 \times 10^{-27} e\, {\rm cm},
\label{xeedmexp}
\end{equation}
where $e=|e|$ is the electric charge. Now, considering our recommended values as 
\begin{eqnarray}
d_\mathrm{a} = 0.510(10) \times 10^{-20} \langle \sigma \rangle C_\mathrm{T} \ \text{e-cm}
\end{eqnarray}
and 
\begin{eqnarray}
d_\mathrm{a}= 0.337(10) \times 10^{-17} \ {S/(e \, \text{fm}^3)} \ \text{e-cm},
\end{eqnarray}
and combining them with the experimental result for EDM, we obtain limits as 
\begin{eqnarray}
 |C_\mathrm{T}| < 4.2 \times 10^{-7}
\end{eqnarray}
and
\begin{eqnarray}
 |S| < 4.2 \times 10^{-10} \ e\, \text{fm}^3.
\end{eqnarray} 

At the hadron level, we have
\begin{eqnarray}
 |\bar{g}_{\pi N N}^{(0)}| &<& 1.2 \times 10^{-9}, \\
 |\bar{g}_{\pi N N}^{(1)}| &<& 1.1 \times 10^{-9}, \\
 |\bar{g}_{\pi N N}^{(2)}| &<& 5.4 \times 10^{-10}
\end{eqnarray}
and 
\begin{eqnarray}
 |d_n| &<& 1.3 \times 10^{-22} \ e\, \text{cm},
\end{eqnarray} 
where we assumed 30\% of nuclear level uncertainty. We do not set a limit for the proton EDM which is affected by large error. When the sensitivity of $^{129}$Xe EDM experiment improves by about three orders of magnitude as expected \cite{Terrano:2021zyh}, the resulting NSM limit together with nuclear structure calculations will give improved limits at the quark-gluon level CP violation.

Using the results from the present study, the final expression for in terms of all possible contributions can be given by
\begin{eqnarray}
d_{\rm Xe} &=& 1.15 \times 10^{-3} d_e \nonumber\\
&& -2.6 \times 10^{-6} d_u +1.0 \times 10^{-5} d_d \nonumber\\
&&+ ( -2 \times 10^{-20} \bar \theta e\, {\rm cm} )\nonumber\\
&&  +2.4 \times 10^{-3} e (\tilde d_d -\tilde d_u) \nonumber\\
&& +
\Bigl(
0.040 C^{eu}_S +0.041 C^{ed}_S \nonumber\\
&& \hspace{1.5em}
-0.29 C^{eu}_P +0.30 C^{ed}_P \nonumber\\
&& \hspace{1.5em}
-0.055 C^{eu}_T +0.22 C^{ed}_T
\Bigr) \times 10^{-20} e\, {\rm cm} , \ \ \ 
\end{eqnarray}
where all elementary level couplings are renormalized at the scale $\mu =1$ TeV.
The experimental upper limit, given by Eq. (\ref{xeedmexp}), is then converted to
\begin{eqnarray}
 |d_e| &<& 1.2 \times 10^{-24}e \, {\rm cm}, \\
 |d_u| &<& 9.0 \times 10^{-22}e \, {\rm cm}, \\
 |d_d| &<& 2.2 \times 10^{-22}e \, {\rm cm}, \\
 |\tilde d_u|, | \tilde d_d| &<& 1.5 \times 10^{-24} {\rm cm}, \\
 |C^{eu}_S| &<& 5.9 \times 10^{-6}, \\
 |C^{ed}_S| &<& 5.7 \times 10^{-6}, \\
 |C^{eu}_P| &<& 8.2 \times 10^{-7}, \\
 |C^{ed}_P| &<& 7.7 \times 10^{-7}, \\
 |C^{eu}_T| &<& 4.2 \times 10^{-6}
\end{eqnarray}
and
\begin{eqnarray}
 |C^{ed}_T| &<& 1.0 \times 10^{-6} .
\end{eqnarray}
This is under the assumption of the dominance of only one P,T-odd interaction.
We also assumed that the quark EDMs, $C^{eq}_S$, $C^{eq}_P$, and $C^{eq}_T$ are affected by 40\% of uncertainty, while the chromo-EDMs by 60\%.

\section{Conclusion}

We have employed relativistic coupled-cluster theory in the linear response approach to estimate the second- and third-order perturbative contributions due to parity and time-reversal symmetry violating interactions to the electric dipole moment of $^{129}$Xe. We have also compared our results with the previously reported values at the random phase approximation, and perform calculation of electric dipole polarizability to verify reliability of our calculations. We observed contrasting trends of correlation contributions in the determination of all these quantities. Especially, determination of third-order perturbative contributions are very sensitive to the contributions from very high-lying $s$ and $p_{1/2}$ orbitals. In addition, we have also performed nuclear calculations using the shell model. Combining atomic results with the latest experimental value of electric dipole moment of $^{129}$Xe, we inferred revised limits of the nuclear Schiff moment and tensor-pseudotensor electron-nucleus coupling coefficient. Using the extracted nuclear Schiff moment with our nuclear calculations, we obtained limits on the pion-nucleon coupling coefficients, and electric dipole moments of a proton and neutron. Further, we used all possible second- and third-order perturbative contributions to express electric dipole moment of $^{129}$Xe in terms of electric dipole moments of electrons and quarks, and parity and time-reversal violating electron-quark tensor-pseudotensor, pseudoscalar-scalar and scalar-pseudoscalar coupling coefficients. 

\section*{Acknowledgement}

BKS acknowledges use of ParamVikram-1000 HPC facility at Physical Research Laboratory (PRL), Ahmedabad to carry out all the atomic calculations. NY was supported by Daiko Foundation. KY used computational resources of Fugaku provided by RIKEN Center for Computational Science through the HPCI System Research Project (Project ID: hp230137). KY was supported by JSPS KAKENHI Grant Numbers 22K14031. 

\appendix*

\section{Matrix}

In the Dirac theory, the orbital wave function of an electron,  $|\phi_a(r)\rangle$, is given by
\begin{equation}
\vert \phi_a (r) \rangle = \frac{1}{r} \begin{pmatrix} P_a(r)\chi_{\kappa_a, m_{j_a}} (\theta,\varphi) \\ i Q_a(r) \chi_{-\kappa_a, m_{j_a}}(\theta,\varphi) \end{pmatrix} ,
\end{equation}
where $P_a(r)$ and $Q_a(r)$ denote the large and small components of the radial part, and the $\chi$'s denote the spin angular parts of each component with relativistic quantum number $\kappa_a$, total angular momentum $j_a$ and its component $m_{j_a}$. 

In terms of these wave functions, the single particle matrix element of the dipole operator $D$ is given by
\begin{eqnarray}
\langle \kappa_a || d ||\kappa_b \rangle = \langle \kappa_a|| C^{(1)}||\kappa_b \rangle \int^{\infty}_{0}dr \left ( P_a P_b +Q_a Q_b \right ) r  , \ \ \ \ \
\label{eqd}
\end{eqnarray}
where $C^{1}$ is the Racah operator of rank 1.

The single particle matrix element of the electron EDM interaction Hamiltonian is given by
\begin{eqnarray}
 \langle j_a || h_k^{d_e} || j_b \rangle =  2 c \sqrt{2j_a +1} \delta_{\kappa_a, -\kappa_b} \nonumber \\ \times \left \{ \tilde{l}_a  (\tilde{l}_a + 1) \int_0^{\infty} dr \frac{ P_a(r) Q_b(r) } {r^2 } +  l_a (l_a+1) \nonumber \right. \\ \left.  \times
 \int_0^{\infty} dr \frac{ Q_a(r) P_b(r)} {r^2} + \frac{dP_a(r)}{dr} \frac{dQ_b(r)}{dr} \nonumber \right. \\ \left. +  \frac{dQ_a(r)}{dr} \frac{dP_b(r)}{dr} \right \} , \ \ \ \ \ \ 
 \label{eqde}
\end{eqnarray}
where $l$ and $\tilde{l}$ are the orbital quantum number of the large and small component of the Dirac wave function respectively.

The single particle matrix elements of the $M1_{hf}$ operator is given by
\begin{eqnarray}
\langle \kappa_a || t^{1}_{hf} ||\kappa_b \rangle =-(\kappa_a+\kappa_b)\langle -\kappa_a|| C^{(1)}||\kappa_b \rangle \nonumber \\
 \times \int^{\infty}_{0}dr \frac{(P_a Q_b +Q_a P_b)} {r^{2}} ,
 \label{m1hf}
\end{eqnarray}
where $\mu_{N}$ is the nuclear magneton and $g_{I}$ is the ratio of nuclear magnetic dipole moment $\mu_{I}$ and $I$.

The single particle reduced matrix element of $h^B(r)$ is given by
\begin{eqnarray}
 \langle j_a || h_k^B || j_b \rangle = \frac{d_e \mu}{2 m_p c} \left \{ -3 \langle -\kappa_a || C^1 || - \kappa_b \rangle \int_R^{\infty} dr \frac{ Q_a(r) P_b(r)} {r^3} \nonumber \right. \\ \left. - 3 \langle \kappa_a || C^1 ||  \kappa_b \rangle \int_R^{\infty} dr \frac{ P_a(r) Q_b(r)} {r^3} - \langle -\kappa_a || \sigma_k || \kappa_b \rangle  \nonumber \right. \\ \left. \times   \int_R^{\infty} dr \frac{ Q_a(r) P_b(r)} {r^3} -  \langle \kappa_a || \sigma_k ||  -\kappa_b \rangle \int_R^{\infty} dr \frac{ P_a(r) Q_b(r)} {r^3} \nonumber \right. \\ \left. + 2 \langle -\kappa_a || \sigma_k || \kappa_b \rangle \int_0^{R } dr \frac{ Q_a(r) P_b(r)} {r^3} \nonumber \right. \\ \left. +  2 \langle \kappa_a || \sigma_k ||  -\kappa_b \rangle \int_0^{R} dr \frac{ P_a(r) Q_b(r)} {r^3} \right \} , \ \ \ \
 \label{eqB}
\end{eqnarray}
where $R$ is the radius of the nucleus.

 The single particle matrix element for the NSM operator is given by
\begin{eqnarray}
 \langle j_a || h_k^{NSM} || j_b \rangle =   \frac{3S}{B} \langle \kappa_a || C_k^{(1)}|| \kappa_b \rangle \nonumber \\ \int_0^{\infty} dr \rho_\mathrm{N}(r) \left ( P_a(r) P_b(r) +  Q_a(r) Q_b(r)  \right ) .
 \label{eqnsm}
\end{eqnarray}

The single particle matrix element of the S-PS interaction is given by
\begin{eqnarray}
 \langle j_a || h_k^{SPs} || j_b \rangle = -  \delta_{\kappa_a, -\kappa_b} \frac{ G_\mathrm{F} C_\mathrm{S}} { \sqrt{2} } A \sqrt{2j_a +1} \nonumber \\  \times \int_0^{\infty} dr ( P_a(r) Q_b(r) + Q_a(r) P_b(r) ) \rho_\mathrm{N}(r)  .
 \label{eqsps}
\end{eqnarray}

The single particle reduced matrix element of Ps-S operator is given by
\begin{eqnarray}
 \langle j_a || h_k^{PsS} || j_b \rangle = -  \frac{ G_\mathrm{F} C_\mathrm{P}} {2 \sqrt{2} m_p c} \langle \bm{\sigma}_\mathrm{N} \rangle \langle \kappa_a || C^{(1)} || \kappa_b \rangle \nonumber \\  \times \int_0^{\infty} dr ( P_a(r) P_b(r) - Q_a(r) Q_b(r) ) \frac{ d\rho_\mathrm{N}(r)} {dr}  .
 \label{eqpss}
\end{eqnarray}

The single particle reduced matrix element of T-Pt operator is given by
\begin{eqnarray}
 \langle j_a || h_k^{TPt} || j_b \rangle = - \sqrt{2} G_\mathrm{F} C_\mathrm{T} \langle \bm{\sigma}_\mathrm{N} \rangle \left [ \langle \kappa_a || \sigma_k || -\kappa_b \rangle \nonumber \right. \\ \left. \times \int_0^{\infty} dr \rho_\mathrm{N}(r) P_a(r) Q_b(r) + \langle - \kappa_a || \sigma_k || \kappa_b \rangle \nonumber \right. \\ \left. \times \int_0^{\infty} dr \rho_\mathrm{N}(r) Q_a(r) P_b(r)  \right ] ,
 \label{eqtpt}
\end{eqnarray}
where $\sigma_k$ is the Pauli spinor for the electrons.

\end{document}